\documentclass[review]{elsarticle}
\usepackage{lineno,hyperref}
\usepackage{booktabs}
\usepackage{comment}
\usepackage{graphicx}
\usepackage{color}
\usepackage{bm}

\begin{document}

\begin{frontmatter}

\title{Alternative way to characterize a q-Gaussian distribution by a robust heavy tail measurement\\}

\author[mymainaddress]{E.L de Santa Helena \corref{mycorrespondingauthor}}
\cortext[mycorrespondingauthor]{Corresponding author. Tel.:+55 7991363772}
\ead{elsh@ufs.br}

\author[mymainaddress]{C. M. Nascimento}

\author[mysecondaryaddress]{G. J. L. Gerhardt}

\address[mymainaddress]{Departamento de F\'{\i}sica, Universidade Federal do Sergipe, Aracaju, Brazil.}
\address[mysecondaryaddress]{Departamento de F\'{\i}sica e Qu\'{\i}mica, Universidade de Caxias do Sul,  Brazil.}

\begin{abstract}

The q-Gaussians are a class of stable distributions which are present in many scientific fields, and
that behave as heavy tailed distributions for an especific range of $q$ values.  
The identification of these values, which are used in the description of systems, is sometimes a hard task. 
In this work  the identification of a q-Gaussian distribution from empirical data was done by a measure of its tail weight using robust statistics. 
Numerical methods were used to generate artificial data, to find out the tail weight -- medcouple, and also to adjust the curve between medcouple and the $q$ value.
We showed that the medcouple value remains unchanged when the calculation is applied to data which have long memory.
A routine was made to calculate the $q$ value and its standard deviation, when applied to empirical data.
It is possible to identify a q-Gaussian by the proposed methods with higher precision than in the
literature for the same data sample, or as precise as found in the literature.
However, in this case, it is required a smaller sample of data.
We hope that this method will be able to open new ways for identifying physical phenomena that belongs to nonextensive frameworks.

\end{abstract}


\end{frontmatter}


\section{Introduction}

It is sometimes a puzzling question to unveil the statistical description related to empirical data coming from physical systems.  
In general, to know the probability distribution function (PDF) related to the system
is not sufficient to precisely describe it, and the information about their correlation relationships are also necessary.
In this sense, in the last decades, many attempts to describing phenomena in several scientific fields were done neither by
their PDF nor by exact correlations, but in terms of 
their asymptotic behaviors. In most cases,
this approach can be justified by observation of a power law asymptotic behavior of the systems' variables. 

Power law distributions are present in many scientific fields, such as
Physics~\cite{P01}, General Science~\cite{P04,P05}, Geophysics~\cite{P02}, Social Sciences~\cite{P06,P07,P08}, Ecology~\cite{P09}, and Economics~\cite{P10}.
However, there are several systems which present a characteristic scale, consequently a power law behavior cannot be observed in them.
Although in both cases the PDFs associated with the dynamical variables are stable, for
the first one
the second moment is not well defined~\cite{P12}, which is a necessary condition for the observation of the power law. See refs.~\cite{P11,P13} for a great discussion concerning this kind of approach and its observation for a diverse range of phenomena.
 
Yet in this line of approach, the description of several phenomena has been done by q-Gaussian 
distributions~\cite{T09}. Although stable, the usual requirement of the independent dynamical variable is not necessary for these distributions~\cite{P14}.  
The q-Gaussian probability density function~\cite{T09}, usually named qPDF, with $q$-mean $\mu_q$ and $q$-variance $\sigma_{q}$ is:
\begin{equation}
\rho(x;\mu_q,\sigma_q)=\alpha\sqrt{\frac{\beta}{\pi}}
\left[ 1+(q-1)\beta(x-\mu_q)^2  \right]^{1/(1-q)}
\label{pdfro}
\end{equation}
where $\beta=[(3-q)\sigma^{2}_{q}]^{-1}$  and
\begin{equation}
\alpha=\left\{
\begin{array}{ll}
\sqrt{1-q} \Gamma(\frac{5-3q}{2-2q}) /  \Gamma(\frac{2-q}{1-q})  & \textrm{if  $q<1$}, \\
1  & \textrm{if  $q=1$}, \\
\sqrt{q-1} \Gamma(\frac{1}{q-1}) / \Gamma(\frac{3-q}{2q-2})  & \textrm{if $1<q<3$}. \\
\end{array}\right.
\end{equation}
Replacing $Z=(q-1)/(3-q)$ and assuming $\mu_q=0,\sigma_q=1$ the standard q-Gaussian can be written as
\begin{equation}
\rho(x;0,1)=\alpha'  \sqrt{ \frac{Z}{\pi}} \left[1+Z x^2  \right]^{-(Z+1)/2}\;,
\label{qpdf}
\end{equation}
where $\alpha'=\alpha/\sqrt{(q-1)}$. In the limit of $q\rightarrow 1$ a qPDF tends to a standard Gaussian distribution. For $q < 1$, 
it is a compact support. When $1 < q < 3$,  it is a heavy tail. In the last case, a power law asymptotic behavior describes well this class of distribution.

A weakness of the treatment of a system's distribution, whose variables are approached according to a non extensive framework with an unknown dynamics, is the difficulty in associating a $q$ value to it.
In most cases, this difficulty arises due to finite data size effects,   
and the curve fitting of the distribution by ordinary methods does not work accurately.
In this paper, we propose an alternative way to identify a q-Gaussian distribution
from empirical data. This identification is done by the medcouple method.

It was demonstrated that the medcouple makes a calculation 
in such a way that the result is 
almost the same as in the case of outliers absence, even in the case when the
data is contaminated with up to $12.5\%$ of them. 
The performance of medcouple and other methods was compared by picking an 
uncontaminated Gaussian distribution as a null hypothesis instead of other alternative hypotheses (non Gaussian distributions). The results pointed out that medcouple is a more conservative among all other tested methods once it rejected much less the null hypothesis~\cite{MMY06}. 
Consequently, picked data from a Gaussian tends to 
be correctly identified by medcouple. In dealing with small size data the medcouple's performance is even 
better, as this approach avoids misidentification through finite size effects. 
Moreover, the method does not demand a finite moments' distribution.

In the theoretical background we presented (i) the medcouple  and (ii) the long memory process definition.
In the following section we describe how the numerical calculations are carried on.
In the results we show how to calculate the $q$ value and its standard deviation from the heavy tail measure. At the same time, we show that the value of medcouple does not change when the PDF is associated to a process of long memory.
In the last section, we discuss the advantage of the proposed method compared to the usual methods of probability distribution fit.
At appendix, we list the $R$\cite{R08} routine that allows to estimate $q$ value and standard deviation from empirical data.

\section{Theoretical Background}
\label{background}
 
\subsection{Medcouple}
The kurtosis is a classical measurement of tail weight of a distribution that is very sensitive to outlying values. Outliers occur in the data set due to measurement errors or contamination and may become more apparent when the sample size
is small. The robust statistics \cite{MMY06} seeks to estimate moments and derived quantities
to nicely fit the bulk of the data when the data contain, or not, outliers.
The median is an example of a robust estimator of the data middle
and a measurement using this estimate should help to identify a heavy tail distribution.
Since we are seeking a method to identify a q-Gaussian 
from the experimental data, we will use the medcouple originally 
introduced by Brys et al. \cite{BHS04} to quantify skewness as follows.
Given a sorted sample $\{x_{1} <...< x_{n}\}$ from a univariate distribution,
the kernel function is defined as:
\begin{equation}
\label{kernel}
h(x_i,x_j)=\frac{(x_j-\tilde{x} )-(\tilde{x} -x_i)}{(x_j-x_i)}
\end{equation}
that measures the distances from $x_i$ and $x_j$ to the median $\tilde{x}$.
Remembering the median definition $\tilde{x}$ applied to this set:
\begin{equation}
\label{defmed}
\textrm{$med(x)$} = \left\{
\begin{array}{ll}
x_{(n+1)/2} & \textrm{if $n$ is odd}  \\
(x_{n/2}+x_{(n/2)+1})/2 & \textrm{if $n$ is even} \\
\end{array}\right.
\end{equation}
the medcouple is defined as:
\begin{equation}
MC=med_{x_{i} \leq \tilde{x}   \leq x_{j}} h(x_{i},x_{j})
\label{kernel_median}
\end{equation}
applied to all pairs that satisfied condition $x_i \leq \tilde{x}   \leq x_j$.
The main feature of this measure is that it is invariant under scale and location changes. For more details, see Brys et al. \cite{BHS04}.
 
The application of medcouple in each side of the distribution leads to two measures which allow quantification of the tail weight \cite{BHS06}.
\begin{equation}
\label{mcl}
LMC=-MC(\{x_{1} <...<\tilde{x} \})
\end{equation}
and
\begin{equation}
\label{mcr}
RMC=MC(\{\tilde{x}<...<x_{n} \} )
\end{equation}
the left and right medcouple.
In the case of symmetric distributions, both measures are equivalent.

\subsection{Long Memory process and Self-Similarity}
A stochastic process caracterized by a probability density function (PDF) $Y_{t}$ ($t$ is the time parameter) is called self-similar when the rescaled PDF $c^{-H}Y_{ct}$ (time scale $ct$ and $c>0$) presents the PDF of the original process \cite{Be94}.

We will consider only $Y_{t}$ self-similar with stationary increments: 
$\left\langle X_{i}\right\rangle=\left\langle Y_{i}-Y_{i-1}\right\rangle=\textrm{constant}$, 
for any time lags $(n>0)$. The covariance between $X_{i}$ and $X_{i+n}$ can be shown to be
\begin{equation}
\label{gamav}
\gamma(n)=\frac{1}{2}\sigma^{2}[ \left|n+1 \right|^{2H}-2\left|n\right|^{2H}+\left|n-1 \right|^{2H}],
\end{equation}
where $H$ is known as the Hurst exponent, and $\sigma^{2}$ is the variance of the increment process $X_{i}=Y_{i}-Y_{i-1}$.
$\gamma(n)$ is non-negative only when $ 0 <H < 1$ and can be seen in this case 
as a legitimate covariance \cite{Sa06}. A long memory process with covariance given by 
eq. (\ref{gamav}) is called fractional Gaussian noise
and the corresponding self-similar process $Y_t$ is called fractional Brownian motion. 

The correlation $\rho(n)=\gamma(n)/\sigma^{2}$ has the asymptotic behavior $\rho(n)=H(2H-1)n^{2H-2}$ 
from which we conclude that $\lim_{n\rightarrow \infty}\rho (n)=0$ when $H<1$. 
When $H=1/2$, the process $X_i$ is uncorrelated because $\rho (n)=0$ for any lag $n\neq 0$.
For $1/2 < H <1$ the process has long range memory since $\sum{\rho (n)}=\infty$ for all $n$,
and for $0 < H <1/2$ the process has short range dependence because $\sum{\rho (n)}=0$.

Samorodnitsky \cite{Sa06} draws attention to the important issue that a
long memory increment process gradually seems to stop showing the stationary behavior 
when the correlation, measured by $H$, is far from a half.
This caveat will be used in a careful choice of a long memory process.

Given the covariance matrix $\Sigma$, 
\begin{equation}
\label{mgamma}
\Sigma=\left(	
\begin{array}{cccc}
\gamma_{0}& \gamma_{1} & \cdots & \gamma_{N-1} \\
\gamma_{1}& \gamma_{0} & \cdots & \gamma_{N-2} \\
\vdots & \vdots& \vdots& \vdots \\
\gamma_{N-1}& \gamma_{N-2} & \cdots & \gamma_{0} \\
\end{array}
\right)
\end{equation}
where the elements of the matrix $\gamma_n$  are obtained from eq. (\ref{gamav}),
we obtain $ L $ by Choleski factorization $\Sigma=LL'$ where $L$ is lower triangular.

For artificially generating a long memory process \cite{Be94,DR91} $\vec{X}_{c}$, we multiply a vector of an independent stationary 
process $\vec{X}$ by a transformation matrix $L$.
\begin{equation}
\label{scorr}
\vec{X}_{c}=L\vec{X}
\end{equation}
To our knowledge, there is no fast algorithm to generate long-memory q-Gaussian noise.

\section{Numerical Calculus}

The box-muler algorithm was implemented as described in \cite{TNT07}
using the Mersenne-Twister algoritm as a random number generator in R\cite{R08}
to generate the q-Gaussian probability density function (qPDF), eq~(\ref{qpdf}).
For any distinct $q$ value, this procedure allowed us to create 
$K=2^{8}$ artificial time series $\vec{X_{k}}(q)$ (replications) 
in three length scales $N=\{2^{13},2^{14},2^{15}\}$.

The Robustbase \cite{RB14} is a robust statistical package that implements the calculation of medcouple
as described in \cite{BHS04}, using a fast algorithm that only needs $O(n \log n)$ time.
The heavy tail measurements, $RMC$ and $LMC$, as described in \cite{BHS06},
was implemented in a straightforward way (see Appendix). 
As we are only dealing with symmetric distributions, the $RMC$ choice was made through a coin toss.
Time series with different lengths $ N $ were used to estimate the $RMC$ standard deviation, 
while only the series with $ N = 2 ^ {14} $ were used in curve fittings by
nonlinear least-squares. We chose a representative set of $q$ values in the range $-1 < Z(q) < 6$
to build up $m(q)$ PDFs of random variable $m_{k}(q)=RMC(\vec{X_{k}}(q)), k=1,2,..K$.
This is the first time it is shown that the medcouple may be used to characterize q-Gaussian distributions of compact support. 

The Choleski factorization is used to generate stationary long memory process.
Since this method is computationally heavy, we need to keep the length of the vector in an acceptable computational size. 
Furthermore, to figure out what the covariance matrix elements (\ref{gamav}) are, 
we use a $H$ value slightly larger than a half to avoid numerical problems
and to ensure a generation of a stationary long memory increment process.

\begin{figure}
\centering
\includegraphics[scale=.42]{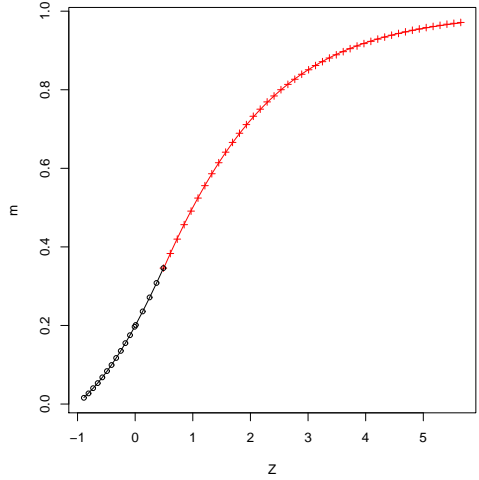}
\caption{\label{fig.1} The mean value $m$ as a function of $Z=(q-1)/(3-q)$. At $Z=0.5$ ($q=5/3$)
there is an inflection point that separates the two parts of the curve fitting eq.~(\ref{maju})  according to the values of table~\ref{tab.1}.}
\end{figure}

\section{Results}

A Shapiro test indicates normality of data sample $\{m_1(q),...,m_K(q)\}$, so we can use
its mean, $m$, as estimate of more probable value. On the other hand, a median is a good estimate when $m(q)$ PDFs has a Poisson shape which occurs at $q < -10$.

\begin{figure}
\centering
\includegraphics[scale=.45]{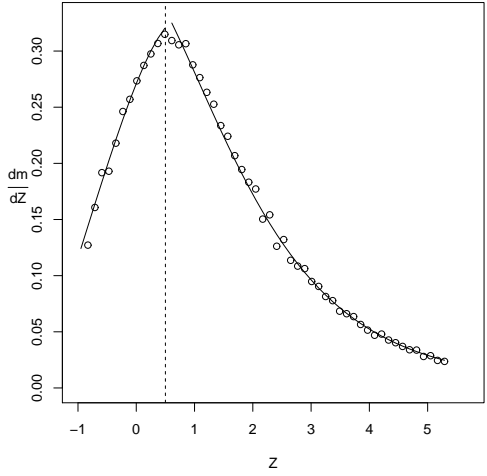}
\caption{The dashed vertical line Z = $ 0.5 $ marks the inflection point ($ q = 5/3 $)
at $dm(Z)/dZ$ which is a function that can be obtained from eq.~(\ref{maju}) and 
table~\ref{tab.1}. The points were obtained by numerical calculus using $dm(Z)/dZ=  (m_{i+1}-m_{i})/(Z_{i+1}-Z_{i})$.
}
\label{fig.2}
\end{figure}

In fig.~\ref{fig.1} we can see $m$ versus $Z$ that gives us an idea of how to make a good curve fitting. In fig.~\ref{fig.2} we can see numerical values of $dm/dZ$ versus $Z$, which allow us to infer the existence of an inflection point
around $Z=0.5 (q=5/3)$. We judge convenient to split the curve fitting in two parts, $-\infty < q < 5/3$ e $5/3 < q < 3$ without any 
difficulty. This procedure aims at providing that the ansatz
\begin{equation}
\label{maju}
m(Z)=\tanh[a+bZ + c Z^2],
\end{equation}
ensure a smooth curve fit that was done with the parameters set $ a, b, c $ listed in table~\ref{tab.1}.
 
\begin{table}[!h]
\caption{Adjust parameters values} 
\begin{center}
\label{tab.1}
\begin{tabular}{ccc}
\toprule

          & $q<5/3$ & $5/3<q<3$ \\
\midrule
$a$  & 0.20177750 &  0.17071450 \\
$b$  & 0.28213917 &  0.38767097  \\
$c$  & 0.08314083 & -0.00837164  \\
\bottomrule
\end{tabular}
\end{center}
\end{table}
Theses curve fitting allows us to calculate an adjusted $q$ value:
\begin{equation}
\label{zaju2}
q(m)=3-\frac{2}{1+Z(m)}
\end{equation}
with $Z(m)$ obtained from eq.~(\ref{maju}).

It was observed that the behavior of the standard deviation of $m(q=1.33)$
as a function of increasing K (replications) decreases quadratically to a value around $ K \approx 2 ^ {8} $ and beyond this point it decreases linearly. Therefore, it is enough to have $ K = 2 ^ {8} $
to obtain a good estimate of the $m$ value standard deviation, $\delta m$
and to assume it as an asymptotic value ($K \rightarrow \infty$)

\begin{figure}
\centering
\includegraphics[scale=.46]{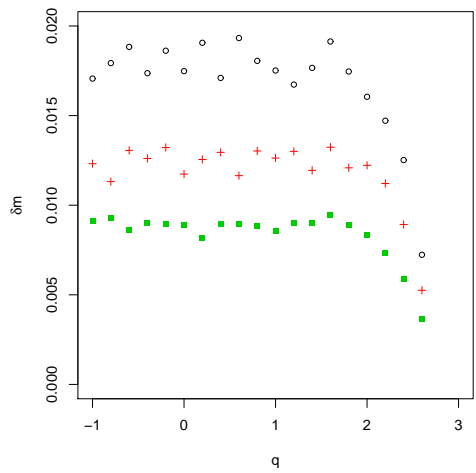}
\caption{In this graph we show the standard deviation, $\delta m$ (see eq.(\ref{delm})) as a function of $ (N,q) $. The points of the upper curve correspond to $ N = 2 ^ {13} $ and those of the lower curve to the values of $ N = 2 ^ {15} $. $\delta m$ is constant only until around $q=5/3$ and
its variability is inversely proportional to $N$ size.} 
\label{fig.3}
\end{figure}

For each one of the $m(q)$ PDFs,  generated from the time series with
different lengths $\{2 ^ {13}, 2 ^ {14}, 2 ^ {15}\}$, the standard deviation $ \delta m $
was calculated  and a graph was drawn as shown in fig.~\ref{fig.3}.
Starting from the different scales we can collapse the data and adjust a relationship between $\delta m $ and $ (N,q) $ as follows:
\begin{equation}
\label{delm}
\delta m \approx \frac{e^{0.5}}{\sqrt{N}} \times \left\{
\begin{array}{ll}
1 & \textrm{if $q\leq5/3$}  \\
0.5^{(q-5/3)} & \textrm{if $q>5/3$} \\
\end{array}\right.
\end{equation}
  
The $q$ value standard deviation $\delta q$ was estimated 
by the usual process of error propagation
\begin{equation}
\label{appro}
\delta q=\left|\frac{dq}{dm}\right|_{m}  \delta m.	
\end{equation}

Since $\delta m $ is fairly constant for $q \leq 5/3$ as can be seen in
fig.~\ref{fig.3} where  the $\delta q $
is governed by the factor $ \left | \frac{dq}{dm} \right | $ that
increases as $ q \rightarrow - \infty $ ($ Z \rightarrow $ -1 ).
This behavior can be infered from eq.~(\ref{zaju2}) and fig.~\ref{fig.2}.
The $\delta q $, beyond $q>5/3$, is governed by the factor $0.5^{(q-5/3)}$, 
since it decreases faster than  $ \left | \frac {dZ} {dm} \right |$ increases.

For symmetric PDFs, we can enhance the $q$ value estimate, including the $LMC$ measurements in calculating the mean and standard deviation of $m$. 
In this case, the accuracy is increased by a $\sqrt{2}$ factor (see fig.~\ref{fig.3}), since it seems as if the size sample, $N$, is doubled. From the results of eq.~(\ref{zaju2},\ref{appro}) and table~\ref{tab.1} we create a routine $R$ (see appendix) to estimate each $q$ value and its standard deviation.

\begin{table}[!h]
\caption{Mean and (standard deviation) of m value 
calculated from uncorrelated and correlated time series}
\begin{center}
\label{tab.2}
\begin{tabular}{cccc}
\toprule
 &  \multicolumn{3}{c}{q}    \\
\cmidrule(lr){2-4}
H & 0 & 1.04 & 1.69 \\
\midrule
0.5 &  .115(.015) & .206(.012) & .358(.011) \\
0.567 & .117(.012) & .205(.012) &  .349(.012)  \\
\bottomrule
\end{tabular}
\end{center}
\end{table}

Many empirical data exhibit long-memory such as financial assets like stocks market returns and currency pairs series. Furthermore, they exhibit  q-Gaussian \cite{RC07,OBT01} behavior.
Therefore, it is useful to verify the behavior of medcouple in these cases.
For this purpose, we use a $2^8$ q-Gaussian noise uncorrelated series of size $N=2^{15} $ to create a long-memory process as described in eq.~(\ref{scorr}). In table~\ref{tab.2} we present some values of $ m $ calculated for some values of $ q $, chosen on a representative range.
The calculation of $ m $ applied to correlated and uncorrelated series has statistically the same value and, therefore, the medcouple is not affected in this case. 
 
\section{Discussion and Conclusions}

Usually, to obtain a reliable fitting of a q-Gaussian distribution to the empirical data, a large amount of data is needed. In geophysics, for example, the author \cite{CPL07} used $ 400.000$ earthquakes to obtain
$q=1.75\pm0.15$ from PDFs of the energy differences and used $10^9$ avalanches to obtain a
PDF of the avalanche size differences caracterized by $q=1.75\pm0.15$
From $2.5\times10^6$ values of temperature fluctuation obtained from WMAP \cite {BTV07}, it was possible to adjust a nonextensive distribution with $q=1.04\pm0.01$.
Liu et al. \cite {LJ08} measured the distribution of position of $10^7$ particles immersed in a plasma and have identified an anomalous diffusion process. After using a low and high energy laser to heat the samples, they obtained distributions of positions fitted with $q = 1.08 \pm 0.01$ e $q = 1.05 \pm 0.014$, respectively.
In economics, the authors \cite{T09,OBT01} discuss how q-Gaussian distributions fit very well 
to empirical distributions of returns SP500 stocks index. For
the empirical return stock volumes from NASDAQ and NYSE they found
$q=\{1.41, 1.44, 1.43\}$ for fitting $10^6$ data points with $\Delta t=\{1,2,3\}\textrm{min}$ time sample.
In these cases, it is reasonable to assume that the uncertainty is in the range of $\pm 0.01$.

We will make some precision comparisons between our results and those above presented.
Choosing the following three $q$ values $\{1.04,1,45,5/3\}$,
we can obtain the standard deviation $\delta q=\pm 0.01$ 
using the data samples with the number of points $\{7\times 10^5, 2 \times 10^5 , 10^5\} $, respectively. 
A smaller sample is sufficient to achieve the same precision when $q>5/3$ (heavy tail distribution) is considered. A possible explanation for this fact can be done analysing the eq.~\ref{kernel}. Picking any pair of points from a sample, getting
the first, $x_i$, near the sample's median, and the second, $x_j$, along the sample's tail,
the yielding values, $h(x_i,x_j)$, will not change neither the median's kernel function (eq.~\ref{kernel_median}) nor its variance. 
Although the medcouple is still better in dealing with small samples with $q \approx 1$ than tradional methods, in this case the gain is less significative than it is for $q$ values larger than $1$. 
Finally, for huge negative $q$ values, the estimator loses its effectiveness, since $dq/dm$ diverges. 
First of all, there are numerical problems for assessing the $RMC$ value
because it is calculated as the median of a set of almost null values (kernel function).
Secondly, because $dq/dZ$ diverges as $ Z \rightarrow $ -1.
 
It is worth calling attention that the q-Gaussian behavior could arise from a normalization process applied to the empirical data \cite{VP09}. This spurious behavior is not observed because medcouple does not need a normalized data set. Furthermore, non-Gaussianity can arise as a finite-size effect in a data analysis \cite{M11}. The medcouple applied to data analyses is less influenced by finite-size effects in comparison with usual methods of curve fitting because it forgets tail values (outliers) at sample 
and it retains data with less influence of the tail, 
only characterizing the sample as Gaussian, when it is true. 
Moreover, the proposed method is not affected if the data have long memory,  
providing an more efficient way to identify empirical distributions instead of replacing those usual ones.
Taking all this into account, the proposed method opens new perspectives for identifying phenomena within nonextensive frameworks.


\section{Appendix}
The $R$ function \footnote{http://200.17.141.35/elsh/qbymc.zip} to calculate the value of $ q $ and its standard deviation is shown below: 

{\footnotesize

library(robustbase)

qbymc=function(x)\{

N=length(x) 

yy=cut((x), c(min(x)-1,median(x), max(x)),label=c(0,1))

mm=data.frame(x,yy) 

vmcl=abs(by(mm[,1],factor(mm[,2]),mc)[2]) 

vmcr=abs(by(mm[,1],factor(mm[,2]),mc)[1]) 

vmct=(vmcl+vmcr)/2

if (vmct $>$ 0.348) j=c(0.1797145,.38767097,-.00837164)

else j=c(0.2017775,.28213917,.08314083)

Z=(-j[2]+sqrt(j[2]${}^\wedge$2-4*j[3]*(j[1]-atanh(vmct))))/(2*j[3])

dqdm=cosh(j[1]+j[2]*Z+j[3]*Z${}^\wedge$2)${}^\wedge$2/((j[2]+2*j[3]*Z)*(1+Z)${}^\wedge$2)

qv=(3*Z+1)/(Z+1)

if (qv $<$ 5/3) dm=exp(.5)/sqrt(N)

else  dm=exp(.5)/sqrt(N)*.5${}^\wedge$(qv-5/3)

dq=dm*dqdm*sqrt(2)

attr(qv, 'names ') $<-$  'Estimate'

attr(dq, 'names' ) $<-$  'Std. Error' 

return(c(qv,dq))

\}
}

\section*{References}

\bibliography{qbymc_ar4}

\end{document}